\documentclass[preprint,floatfix] {revtex4} 
 
\usepackage{graphicx}
\usepackage{subfigure}
\begin{document}

\title{Bound state spectra of the 3D rational potential}
\author{Amlan K. Roy}
\altaffiliation{Corresponding author. Email: akroy@chem.ucla.edu. Present address: Department of
Chemistry, University of Kansas, Lawrence, KS, 66045, USA.}
\affiliation{Department of Chemistry and Biochemistry, University of California, 
Los Angeles, CA, 90095-1569, USA}

\author{Abraham F. Jalbout}
\affiliation{Institute of Chemistry, National Autonomous University of Mexico,
Mexico City, Mexico}

\author{Emil I. Proynov}
\affiliation{Q-Chem Inc., 5001 Baum Blvd., Pittsburgh, PA 15213, USA}

\begin{abstract}
We present bound state spectra of the 3D rational potential, $V(r)=r^2 + \lambda r^2/(1+gr^2)$, $g>0$,
by means of the generalized pseudospectral method. All the thirty states corresponding to 
$n$=0--9 are considered for the first time for a broad range of coupling parameters. These results 
surpass the accuracy of \emph{all} other existing calculations published so far except the 
finite-difference method, which yields similar accuracy as ours. Variation of energies and radial
distribution functions is followed with respect to the interaction parameters. Special emphasis has 
been laid on \emph{higher} excitations and \emph{negative} values of the interaction, where relatively
less work has been reported. The energy sequence is found to be different for positive and negative
interaction; numerically following a mirror-image relationship \emph{usually}, if not always. 
Additionally, twenty energy splittings arising from certain levels belonging to $n$=0--9 are 
systematically studied as functions of the potential parameters. Several new states (including the 
higher ones) are presented.
\end{abstract}
\maketitle

\section{Introduction}
Rational (also termed as non-polynomial oscillator, NPO) potentials are one of the most widely 
studied important model systems in quantum mechanics. The objective of this article is to investigate 
such potentials of the form,
\begin{equation}
V(r)=r^2+\frac{\lambda r^2}{1+gr^2}; \ \ g>0
\end{equation}
The Schr\"odinger equation (SE) can not be solved exactly in general and an 
impressive amount of theoretical works exist employing numerous formalisms as evident from a vast 
amount of literature \cite{haken70}-\cite{saad06}. It has significance in nonlinear Lagrangian methods 
in quantum field theory, laser physics, nonlinear optics, particle physics, etc. 
\cite{haken70,biswas73,whitehead82}. Obviously for $\lambda\!=\!0$ or $\lambda\!=\!g=\!0$ or 
$g\!<<\! \lambda$ or large $g$, its solution behaves as the harmonic oscillator. The potential 
in 1D has attracted enormous attention from a large number of authors for nearly three decades 
employing several powerful tools such as variational method, perturbation theory (PT), 
semi-numerical or purely numerical technique. \emph{Exact} analytical solutions of ground- and 
some excited-state eigenvalues and eigenfunctions in 1D are possible provided $g$ and $\lambda$ 
are related in certain specific ways \cite{varma81,flessas82,whitehead82,chaudhuri83,roy87,roy88a}. 
Some of the most important works are chronologically: Ritz variational principle using Hermite 
polynomial basis \cite{mitra78}, variational and PT calculations \cite{bessis80}, PT using [6,6] 
Pade approximation and hypervirial theorem \cite{lai82}, a variety of finite-difference approaches 
of different flavour \cite{fack85,witwit92,witwit96}, algebraic PT \cite{fack86}, combined 
supersymmetry and WKB approach \cite{roy88a}, continued fraction method \cite{scherrer88}, analytic 
continuation method \cite{hodgson88}, PT with mixed hypervirial and Hellmann-Feynmann theory 
\cite{witwit92a}, Hill-determinant method \cite{agrawal93}, variational bounds via Rayleigh-Ritz 
theorem \cite{stubbins95}, a quadrature discretization technique \cite{chen98} and purely numerical 
calculation \cite{ishikawa02}, etc. Some of these deal with only 1D and others for both 1D plus 3D 
and/or N dimensions.

In parallel to the 1D case, 3D NPO has also been treated in significant detail in the past and 
recent years, and the interest continues to grow. \emph{Exact} solutions were reported for some 
special cases of $g$ and $\lambda$ using SUSY \cite{adhikari91}, generalized harmonic oscillator 
form \cite{znojil83}, etc. However since these are of limited applicability, several methodologies 
were attempted. Some of the promising ones are: shifted 1/N expansion \cite{varshni87,roy88}, 
bounds from eigenvalue moments method \cite{handy93}, hypervirial calculation \cite{witwit91}, 
bounds through the envelope method \cite{saad06}. However as clear, 3D NPO remains relatively less 
explored as compared to its 1D counterpart. Amongst these, satisfactory results for \emph{arbitrary} 
combination of potential parameters of a \emph{general} state can be obtained only by a very few. 
Thus while for 1D case, more than 10 decimal-place accuracy eigenvalues were reported in a number 
of papers (for example \cite{fack86,hodgson88,witwit92a,agrawal93,stubbins95,witwit96,chen98}, to 
our knowledge, only a few of the mentioned works \cite{witwit91,handy93,saad06} are capable of 
delivering such results for 3D NPO satisfying {\em all} these criteria. Moreover two of these works 
in \cite{handy93} and \cite{saad06} deal mostly with estimate of the bounds, not the direct values. 
Thus there is a scarcity of good-quality results for 3D NPO as compared to that of 1D. It may be 
mentioned here that with the rare exception of \cite{varshni87}, practically all works on 3D NPO 
have focused on (+)ve $\lambda$, even though it has been known for long time that $\lambda \!<\! 0$ 
also offers well-behaved solutions provided $g\!\!>$0. Negative $\lambda$ has been examined in 1D 
in detail \cite{agrawal93,stubbins95}, however.

In recent years, generalized pseudospectral (GPS) method has witnessed great success and been 
proved to be a powerful, viable alternative for accurate and efficient treatment of a variety of 
potentials in quantum mechanics, such 
as the spiked harmonic oscillator, the power-law, the logarithmic, the Hulthen, Yukawa, etc., as 
well as both static, dynamic properties of many-electron systems including multiply excited 
high-lying Rydberg states of atoms \cite{roy02}-\cite{sen06}. Successful applications have also 
been made for the 3D spherical confinement studies of the isotropic harmonic oscillator, the H 
atom and the Davidson oscillator \cite{sen06}. A review of the GPS method in the context of atomic 
and molecular calculations is available in \cite{chu05}. For some of the problems such as those in refs. 
\cite{roy04,roy04b,roy05,roy05a}, this has offered results either quite comparable to those
obtained from the best available methods in the literature 
or has surpassed the existing calculations. Another attractive feature is that it was very 
successful to furnish equally high-quality results for both lower and \emph{higher} states, as well 
as for weaker and \emph{stronger} interaction, which was not possible for many of the commonly used 
methodologies (see, for example \cite{roy05a}). However, to our knowledge, the GPS method has not been 
applied in the 
context of NPO so far. Hence our first objective is to extend its scope and validity in the realm 
of this heavily studied important potential. As already hinted, we will treat both $\lambda\!>\!0$ 
and $\lambda\!<\!0$ with special attention to the higher-lying states. To this end, calculated 
energies are presented for a wide range of $g$, $\lambda$. Then the changes in energy are monitored with 
respect to the parameters. Secondly we explore the possible patterns in energy orderings by systematically 
performing all the thirty states with varying $n_r$ and $l$ quantum numbers lying within $n$=0--9 for 
both (+)ve and ($-$)ve $\lambda$. Besides, it is known that the characteristic degeneracy of certain 
energy levels of the 3D harmonic oscillator is removed in certain situations (in the limit of 
$\lambda \! \rightarrow \!0$ or $g\! <<\! \lambda$, for example); first four of such splittings have 
been discussed at length 
for (+)ve $\lambda$ \cite{varshni87,scherrer88} corresponding to $n$=0--4. Here we extend such an
investigation for all the twenty differences possible within the range of $n=0-9$ for both (+)ve and 
($-$)ve $\lambda$ as functions of the coupling parameters. Comparison 
with literature data has been made wherever possible. Section II gives a brief account of the method 
of calculation. Discussion on the results are given in Section III while a few concluding remarks 
are made in Section IV.

\section{Method of calculation}
\label{sec:method}
This section summarizes the essential details of GPS formalism to solve the nonrelativistic radial 
SE of a single-particle Hamiltonian. Relevant details have been presented elsewhere 
(\cite{roy02}-\cite{sen06} and references therein); thus not repeated. Unless otherwise mentioned, 
atomic units are employed throughout this article. 

The desired SE can be written in following form, 
\begin{equation}
\left[-\frac{1}{2} \ \frac{\mathrm{d^2}}{\mathrm{d}r^2} + \frac{\ell (\ell+1)} {2r^2}
+v(r) \right] \psi_{n,\ell}(r) = E_{n,\ell}\ \psi_{n,\ell}(r)
\end{equation}
where $v(r)=V(r)/2$. A 1/2 factor is introduced for sake of consistency with the literature.

The key step is to approximate a function $f(x)$ defined in the interval $x \in [-1,1]$ by an N-th 
order polynomial $f_N(x)$ such that, 
\begin{equation}
f(x) \cong f_N(x) = \sum_{j=0}^{N} f(x_j)\ g_j(x),
\end{equation}
ensuring that the approximation is \emph {exact} at the \emph {collocation points} $x_j$, i.e.,
$f_N(x_j)\!=\!f(x_j).$
In the Legendre pseudospectral method used here, $x_0\!\!=\!\!-1$, $x_N\!\!=\!\!1$, while 
$x_j\!(j=1,\ldots,N-1)$ are obtained from the roots of first derivative of the Legendre polynomial 
$P_N(x)$ with respect to $x$, i.e., $P'_N(x_j)\!=\! 0.$ The cardinal functions $g_j(x)$ given by,
\begin{equation}
g_j(x) = -\frac{1}{N(N+1)P_N(x_j)}\ \  \frac{(1-x^2)\ P'_N(x)}{x-x_j},
\end{equation}
have the unique property, $g_j(x_{j'})\!=\!\delta_{j'j}$. Now (i) mapping the semi-infinite domain 
$r\! \in \![0,\infty]$ onto the finite domain $x\!\!\in\!\![-1,1]$ by the transformation 
$r\!\!=\!\!r(x)$, (ii) using an algebraic nonlinear mapping,
\begin{equation}
r=r(x)=L\ \ \frac{1+x}{1-x+\alpha},
\end{equation}
(iii) followed by a symmetrization procedure leads to the following equation
\begin{widetext}
\begin{equation}
\sum_{j=0}^N \left[ -\frac{1}{2} D^{(2)}_{j'j} + \delta_{j'j} \ v(r(x_j))
+\delta_{j'j}\ v_m(r(x_j))\right] A_j = EA_{j'},\ \ \ \ j=1,\ldots,N-1,
\end{equation}
\end{widetext}
where $A_j \!\! =\!\! \left[ r'(x_j)\right]^{1/2} \psi(r(x_j))\ \left[ P_N(x_j)\right]^{-1}$; 
symmetrized second derivatives of the cardinal function, $D^{(2)}_{j'j}$ are given in the references 
\cite{roy02}-\cite{roy05a}.

By performing a series of test calculations, a consistent set of parameters were chosen 
($\alpha$=25, N=200) which produced ``stable'' converged results. These were used for all the 
calculations reported in this work. Usually $r_{max}$ was set at 150, but for higher excitations larger 
values up to 300 a.u. was used (see tables and Section III for details). The energies are given only 
up to the precision that maintained stability and are {\em truncated} rather than {\em rounded-off}. 

\begingroup
\squeezetable
\begin{table}
\caption {\label{tab:table1}Calculated lowest ($n_r=0$) eigenvalues E (times 2 in a.u.) of the 
3D NPO for several $g$ and $\lambda$ along with literature data for $\ell=0,1,2$. Asterisks denote 
exact analytical values \cite{saad06,roy88}. $r_{max}=150$ a.u.} 
\begin{ruledtabular}
\begin{tabular}{crrrrrrrrr}
        &  &     & \multicolumn{2}{c}{ E}  &  &  & & \multicolumn{2}{c}{ E}  \\
\cline{4-5}  \cline{9-10}
$\ell$ & $g$ & $\lambda$ & This work  & Ref.   &  $\ell$ & $g$  & $\lambda$  & This work & Ref.      
\\  \hline
 0   & 0.1  & $-$0.46   & 2.4000000000000     & 2.400000000000\footnotemark[1],2.4*   &
 1   & 0.1  & $-$0.5    & 3.9999999999999     & 4.000116\footnotemark[2],4*          \\
 0   & 1    & $-$10     & $-$3.000000000000   & $-$3.000000000000\footnotemark[1],$-$3*  &
 1   & 0.01 & $-$0.041  & 4.9000000000000     & 4.899974\footnotemark[2],4.9*        \\
 0   & 10   & $-$640    & $-$57.000000000000  & $-$57.000000000000\footnotemark[1],$-$57*  &
 2   & 0.1  & $-$0.54   & 5.5999999999999     & 5.600000000000\footnotemark[1],    \\
     &      &           &                     &                                           &
     &      &           &                     & 5.599965\footnotemark[2],5.6*       \\
 2   & 1    & $-$18     & $-$6.999999999999   & $-$7.000000000000\footnotemark[1],$-$7*   & 
 2   & 10   & $-$1440   & $-$133.000000000000 & $-$133.000000000000\footnotemark[1],    \\
     &      &           &                     &                                           &
     &      &           &                     & $-$133*       \\
\end{tabular}
\end{ruledtabular}
\footnotetext[1]{Variational calculation \cite{saad06}.}
\footnotetext[2]{Shifted 1/N expansion result \cite{roy88}.}
\end{table}
\endgroup

\section{Results and Discussion}
At first, we examine the convergence of our calculated energy eigenvalues of the 3D NPO. Table I 
compares a number 
of levels for particular values of $g$, $\lambda$ which offer \emph{exact} analytical results 
(denoted by asterisks). These are available only for $\lambda\!\!<\!\!0$ and given for the lowest 
states ($n_r\!\!=\!\!0$) having $l\!\!=\!\!0,1,2$. Note that all energies in this table and 
throughout the article give E multiplied by a 2 factor, for easy comparison with literature. In all 
cases, 12--13 decimal-place accuracy is easily obtained. The variational calculations of 
\cite{saad06} employ a Gol'dman and Krivchenkov Hamiltonian as a solvable model and there is excellent
agreement of our results with theirs for all cases. Some of these were reported long times ago through 
shifted 1/N expansion \cite{roy88} and quoted appropriately. The present results clearly 
outperform them. As a second test, a few representative calculations were performed with 
$\lambda$=0 for several low and high $l$. In all occasions, expectedly harmonic oscillator 
eigenvalues were recovered promptly to at least 13th place of decimal and hence omitted. This amply 
demonstrates the power and authenticity of GPS method in the present context.

\begingroup
\squeezetable
\begin{table}
\caption {\label{tab:table2}First two ($n_r=0,1$) eigenvalues E (times 2 in a.u.) corresponding to 
$l=0-3$ of 3D NPO for select $g$ and $\lambda$ along with the literature data. $r_{max}=150$ a.u.} 
\begin{ruledtabular}
\begin{tabular}{ccccrl}
    &   &     &           & \multicolumn{2}{c}{E} \\
\cline{5-6} 
$n_r$ &  $\ell$ & $g$ & $\lambda$ & This work    & Reference  \\ \hline
0    &  0   &  0.1    &  0.1  & 3.120081864016  & 3.1200\footnotemark[1]   \\         
1    &      &         &       & 7.231009980656  & 7.2312\footnotemark[1]   \\         
0    &  1   &         &       & 5.186373002931  & 5.1863730029314$<$E$<$5.1863730029316\footnotemark[2],
          5.1863730029314\footnotemark[3],5.186338\footnotemark[4],5.1864\footnotemark[1]    \\
1    &      &         &       & 9.276541985488  & 9.276635\footnotemark[4],9.2766\footnotemark[1]  \\
0    &  2   &         &       & 7.2439618404219 & 7.2439618404138$<$E$<$7.2439618404260\footnotemark[2],
          7.2439618404189\footnotemark[3],7.243927\footnotemark[4],7.244\footnotemark[1]  \\
1    &      &         &       & 11.317997742355 & 11.258\footnotemark[1]          \\
0    &  3   &         &       & 9.294359110874  & 9.29435911086337$<$E$<$9.29435911088159\footnotemark[2],
          9.2943591108746\footnotemark[3],9.2944\footnotemark[1]            \\
1    &      &         &       & 13.355727291254 &                                             \\
0    &  0   &  1      &  1    & 3.507388348905  & 3.50738835\footnotemark[3],
                                3.50738835\footnotemark[5],3.5122\footnotemark[1] \\
1    &      &         &       & 7.648201241719  & 7.64820124\footnotemark[3],
                                7.64820124\footnotemark[5],7.6252\footnotemark[1]        \\
0    &  1   &         &       & 5.651393306756  & 5.6503$<$E$<$5.6521\footnotemark[2], 
          5.651393317250\footnotemark[3],5.652112\footnotemark[4],5.6522\footnotemark[1]     \\
1    &      &         &       & 9.713754138848  & 9.705584\footnotemark[4],9.7056\footnotemark[1] \\
0    &  2   &         &       & 7.734828038042  & 7.734$<$E$<$7.736\footnotemark[2], 
          7.734828042923\footnotemark[3],7.734778\footnotemark[4],7.7348\footnotemark[1]        \\
1    &      &         &       & 11.76582837889  & 11.7630\footnotemark[1]                      \\
0    &   3  &         &       & 9.787669778003  & 9.7875$<$E$<$9.7881\footnotemark[2], 
          9.787669778509\footnotemark[3],9.7876\footnotemark[1]                         \\
1    &      &         &       & 13.804700633187 &                       \\
%%%0    &   0  &  10     & 10    & 3.879036830882  & 3.879037\footnotemark[6],3.5732\footnotemark[1] \\
%%%1    &      &         &       & 7.903154159824  & 7.903154\footnotemark[6],7.0990\footnotemark[1] \\
%%%0    &   1  &         &       & 5.940860388291  & 5.940860\footnotemark[6],5.940689\footnotemark[4],
%%%                                                  5.9408\footnotemark[1]  \\ 
%%%1    &      &         &       & 9.944897854236  & 9.944898\footnotemark[6],9.950350\footnotemark[4],
%%%                                                  9.9504\footnotemark[1]   \\
%%%0    &   2  &         &       & 7.962229649097  & 7.962230\footnotemark[6],14.089958\footnotemark[4],
%%%                                                  7.9622\footnotemark[1]         \\ 
%%%1    &      &         &       & 11.963342886765 & 11.963343\footnotemark[6],11.9654\footnotemark[1]  \\
%%%0    &   3  &         &       & 9.9724553529293 & 9.972455\footnotemark[6],9.9724\footnotemark[1]    \\
%%%1    &      &         &       & 13.972886840962 &                                          \\ 
0    &  0   & 100     & 100   & 3.983098339488  & 3.983098\footnotemark[6],3.9844\footnotemark[1] \\
1    &      &         &       & 7.984443523273  & 7.984444\footnotemark[6],7.9910\footnotemark[1] \\
0    &  1   &         &       & 5.993438790399  & $-<$E$<$6.389\footnotemark[2],
5.993438873366\footnotemark[3],5.993439\footnotemark[6],5.993565\footnotemark[4],5.9936\footnotemark[1] \\
1    &      &         &       & 9.993516159965  & 9.993516\footnotemark[6],9.994694\footnotemark[4],
         9.9946\footnotemark[1]                                                              \\
0    &  2   &         &       & 7.996024670900  & 7.9947$<$E$<$8.037800\footnotemark[2],
7.996024673021\footnotemark[3],7.996025\footnotemark[6],7.996048\footnotemark[4],7.9960\footnotemark[1]  \\
1    &      &         &       & 11.996039234884 & 11.996039\footnotemark[6],11.9964\footnotemark[1]      \\ 
0    &  3   &         &       & 9.997153638476  & 9.9969$<$E$<$10.0113\footnotemark[2],
9.997153638602\footnotemark[3],9.997145\footnotemark[6],9.9972\footnotemark[1]                           \\
1    &      &         &       & 13.99715862578  &                                      \\
\end{tabular}
\end{ruledtabular}
\footnotemark[1]{Shifted 1/N expansion result \cite{varshni87}.}
\footnotetext[2]{Lower and upper bounds from eigenvalue moment method \cite{handy93}.}
\footnotetext[3]{Variational upper bounds \cite{saad06}.}
\footnotetext[4]{Shifted 1/N expansion result \cite{roy88}.}
\footnotemark[5]{Quadratic discretization result \cite{chen98}.}
\footnotemark[6]{Continued fraction result \cite{scherrer88}.}
\end{table}
\endgroup

Now we are ready to present our central result in Table II. The calculated eigenvalues are compared 
with the best 
existing literature data for three representative pairs of ($g$, $\lambda$), \emph{viz.}, (0.1,0.1), 
(1,1) and (100,100), denoted respectively as (a)--(c), covering sufficiently broad ranges of 
interaction. For each of these sets, lowest two values ($n_r$=0,1) belonging to $l$=0,1,2,3 are given. 
Bounds were estimated from the eigenvalue moment method \cite{handy93} for the first states belonging 
to $l\!\ge$1 of all them. As seen, for smaller values of $g$ and $\lambda$ in (a), the bounds are quite 
good; but as one passes to (c) for larger values, they 
deteriorate quite badly. Significantly improved upper bounds for same states ((b) and (c)) 
have been published recently \cite{saad06}. For the highest state considered ($n_r$=1,$l$=3) of each 
pair (a)--(c), no results could be found for comparison. The shifted 1/N expansion results are reported 
for all of these states except the highest ones. The first seven states of (c) have also been studied by a 
continued fraction method \cite{scherrer88} with modest accuracy. Clearly our results turn out to be 
the most accurate direct estimates for all of these cases. Now Fig. 1 displays the changes in 
ground-state energy (times 2) against $g$ for seven fixed values of $\lambda$ ($-$100,$-$50,$-$10,1,100, 
200) in (a) and $\lambda$ for five fixed values of $g$ (0.1,0.5,1,10,100) in (b). Similar plots hold for 
higher energy levels. For a fixed $\lambda$, with increase in $g$, eigenvalues decrease steadily 
approaching those of the harmonic oscillator asymptotically for large $g$, as expected, and a larger 
$\lambda$ shows this behavior with greater magnitude. Correspondingly opposite trend is noticed for 
negative $\lambda$. In (b), we see that the changes against $\lambda$ are more prominent for smaller 
$g$ and as $g$ increases, again eigenvalues approaching those of the harmonic oscillator.

\begin{figure}
\begin{minipage}[c]{0.40\textwidth}
\centering
\includegraphics[scale=0.38]{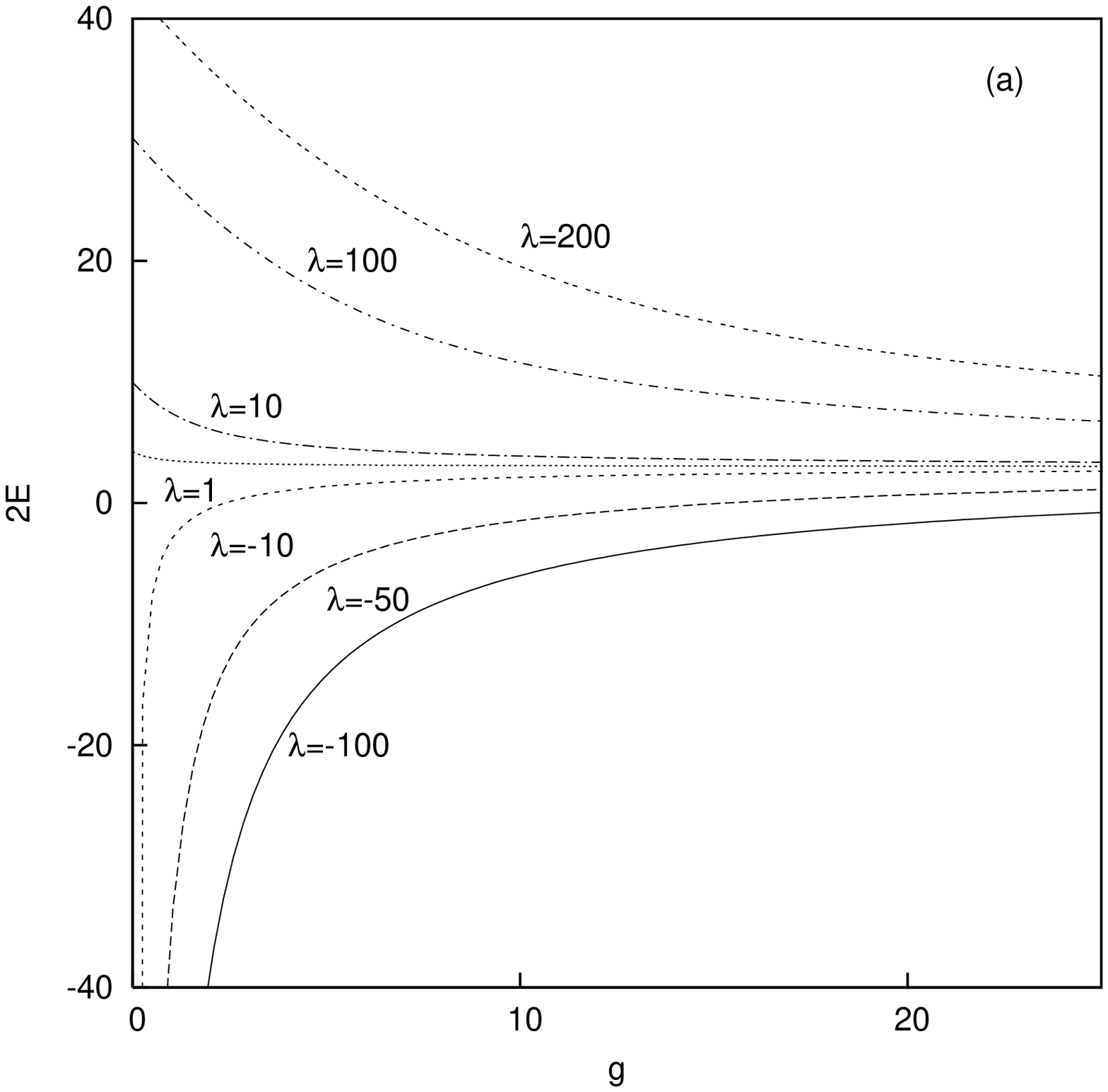}
\end{minipage}%
\hspace{0.3in}
\begin{minipage}[c]{0.40\textwidth}
\centering
\includegraphics[scale=0.38]{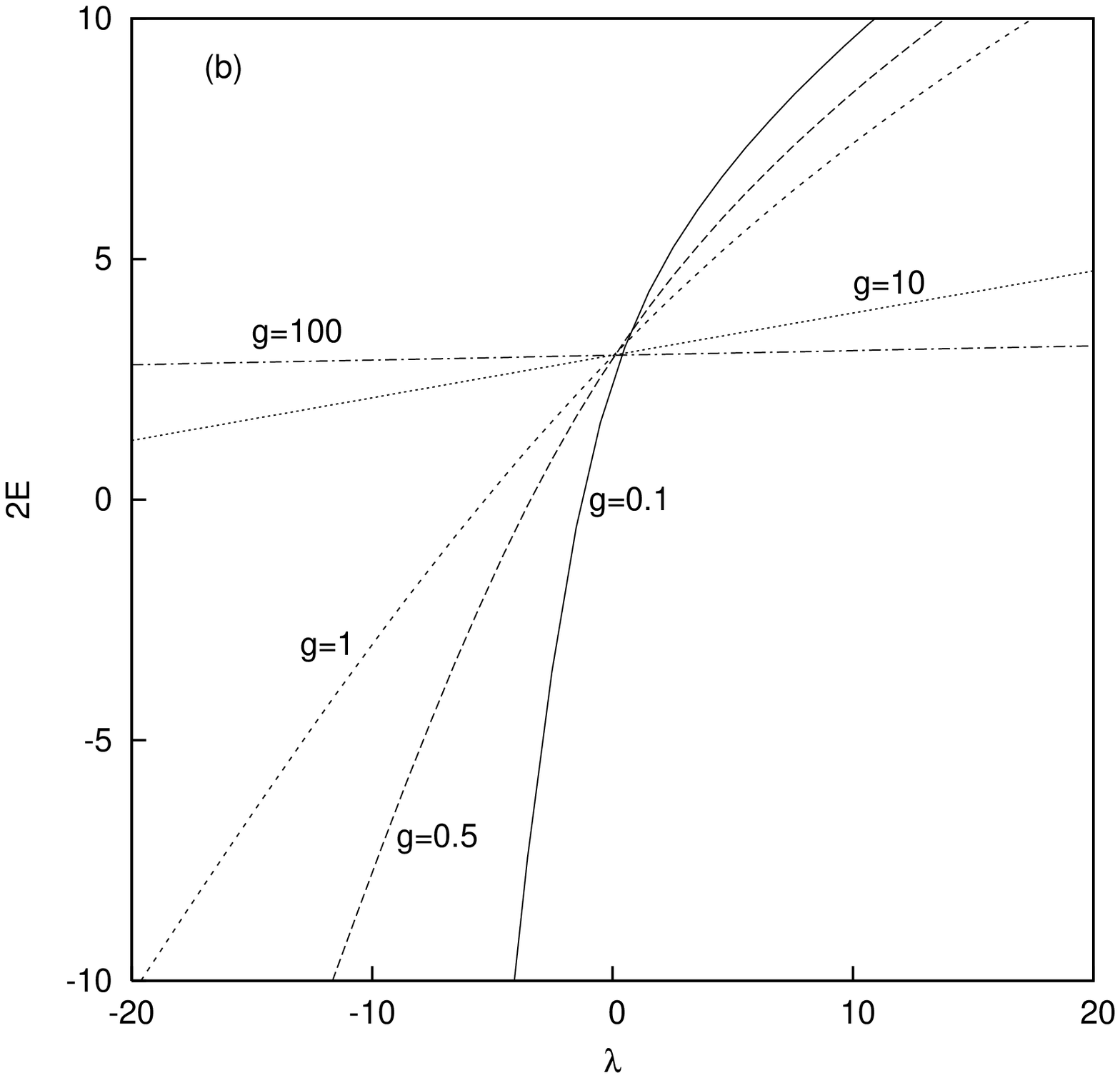}
\end{minipage}%
\caption{Variation of the ground-state energy (2E) of 3D NPO with respect to (a) $g$ for fixed 
$\lambda$, (b) $\lambda$ for fixed $g$.}
\end{figure}

\begingroup
\squeezetable
\begin{table}
\caption {\label{tab:table3}High-lying eigenvalues E (times 2 in a.u.) of the 3D NPO potential for 
some selected $g$ and $\lambda$ along with the literature data. $r_{max}=300$ a.u.} 
\begin{ruledtabular}
\begin{tabular}{cccllccll}
    &   &   &  \multicolumn{2}{c}{E}  &    &  & \multicolumn{2}{c}{E} \\
\cline{4-5}  \cline{8-9} 
$\ell$  & $g$ &  $\lambda$ & This work & Ref. \cite{witwit91} &  $g$ & $\lambda$ & 
                             This work & Ref. \cite{witwit91}    \\ \hline
 10  & 0.1 & 200   & 311.860880892760\footnotemark[1]   & 311.860880892760  
     & 0.2 & 1000  & 699.105622574512\footnotemark[2]   & 699.105622574512  \\
     &     &       & 361.308788624591   & 361.308788624592  &     &       & 811.026062535266   & 811.026062535263  \\ 
     &     &       & 409.563564997653   & 409.563564997653  &     &       & 920.708010354176   & 920.708010354177  \\
     &     &       & 456.909323126874   & 456.909323126874  &     &       & 1028.176013499333  & 1028.176013499331 \\
     &     &       & 503.090316800161   &                   &     &       & 1135.454829499741  &                   \\ 
 20  & 0.5 & 500   & 914.366310994354\footnotemark[3]   & 914.36631099435   
     & 0.1 & 1000  & 1312.25167480938\footnotemark[4]   & 1312.251674809389 \\ 
     &     &       & 990.80662152100    & 990.80662152100   &     &       & 1425.58600163865   & 1425.586001638665 \\ 
     &     &       &1066.14830339382    & 1066.14830339382  &     &       & 1537.79045903952   & 1537.790459039524 \\ 
     &     &       &1140.40019428154    & 1140.40019428154  &     &       & 1648.87112288517   & 1648.871122885173 \\ 
     &     &       &1213.57118403531    &                   &     &       & 1758.83409624488   &                   \\
\end{tabular}
\end{ruledtabular}
\footnotetext[1] {Lower and upper bounds: 311.8601371$<$E$<$311.8616266 \cite{handy93}.}
\footnotetext[2] {Lower and upper bounds: 699.10424$<$E$<$699.109092 \cite{handy93}.}     
\footnotetext[3] {Lower and upper bounds: 914.36540$<$E$<$914.36851 \cite{handy93}.}     
\footnotetext[4] {Lower and upper bounds: 1312.25006$<$E$<$1312.25333 \cite{handy93}.}     
\end{table}
\endgroup

For further test, in Table III, we show some specimen results of the 3D NPO for some higher excited
states. For this we choose two sets of ($g$,$\lambda$); for each of them first five states belonging to 
$l$=10,20 are considered. In the literature, very few results are available for \emph{large} $\lambda$ 
as this often creates problems and this
motivates us to include higher $\lambda$. The high-lying excited states are diffuse and extend 
over a large spatial region. Thus to incorporate these long-range contributions properly, larger 
values (300 a.u.) of $r_{max}$ needed in these cases. Lower and upper bounds for the lowest state of 
each set were reported \cite{handy93} and current results are in consonance with those estimates. 
First four states were also calculated by finite-difference method \cite{witwit91} and as seen, GPS 
results virtually coincide with those accurate values. In many cases, these two offer completely 
identical energies; for others they differ in the last decimal place. For $n_r$=4, no reference could be 
found.

\begingroup
\squeezetable
\begin{table}
\caption {\label{tab:table4}Calculated eigenvalues E (times 2 in a.u.) of 3D NPO for four pairs 
of $g$ and $\lambda$ in parentheses along with literature data for all the states corresponding 
to $n$=0-9. $r_{max}=150$ a.u.}
\begin{ruledtabular}
\begin{tabular}{ccrrccrr}
  & $n$ & (1000,0.1) & (10,1000)\footnotemark[1]$^,$\footnotemark[2] &  &  $n$ 
  & (0.1,$-$1)\footnotemark[3]                                       & (1,$-$100) \\
\cline{3-4} \cline{7-8}
$1s$ & 0 & 3.00009981081 &  64.8250831107 & $1s$ & 0 & 1.5624896124 & $-$79.086464632     \\
$1p$ & 1 & 5.00009993345 &  89.1234516907 & $1p$ & 1 & 2.8620942889 & $-$78.861621366     \\
$2s$ & 2 & 7.00009981589 &  94.8759670627 & $1d$ & 2 & 4.2817900651 & $-$78.421433572     \\
$1d$ &   & 7.00009996001 & 100.703995819  & $2s$ &   & 4.4940422546 & $-$75.296619703     \\
$2p$ & 3 & 9.00009993351 & 101.225823801  & $1f$ & 3 & 5.7883629695 & $-$77.782769536     \\
$1f$ &   & 9.00009997141 & 105.507579307  & $2p$ &   & 6.0427339658 & $-$75.063364459     \\
$3s$ & 4 & 11.0000998195 & 103.184601259  & $1g$ & 4 & 7.3613625227 & $-$76.966703116     \\
$2d$ &   & 11.0000999599 & 105.945405717  & $2d$ &   & 7.6312350478 & $-$74.608167024     \\
$1g$ &   & 11.0000999777 & 108.527833520  & $3s$ &   & 7.7681684221 & $-$71.510476611     \\
$3p$ & 5 & 13.0000999335 & 106.763936963  & $1h$ & 5 & 8.9870011416 & $-$75.995560432     \\
$2f$ &   & 13.0000999713 & 109.683699997  & $2f$ &   & 9.2593659985 & $-$73.950607078     \\ 
$1h$ &   & 13.0000999817 & 111.060023667  & $3p$ &   & 9.4398078008 & $-$71.267997180     \\
%%%$4s$ & 6 & 15.0000998225 & 108.453540747  & $1i$ & 6 & 10.655422039 & $-$74.890837904     \\
%%%$3d$ &   & 15.0000999599 & 110.463776823  & $2g$ &   & 10.923492550 & $-$73.114325497     \\
%%%$2g$ &   & 15.0000999776 & 112.575161928  & $3d$ &   & 11.127177164 & $-$70.796578930     \\
%%%$1i$ &   & 15.0000999845 & 113.394170892  & $4s$ &   & 11.225668905 & $-$67.728270387     \\
%%%$4p$ & 7 & 17.0000999336 & 111.484956198  & $1j$ & 7 & 12.359282590 & $-$73.672049474     \\
%%%$3f$ &   & 17.0000999713 & 113.817393964  & $2h$ &   & 12.619451390 & $-$72.123507544     \\
%%%$2h$ &   & 17.0000999817 & 115.079578535  & $3f$ &   & 12.834220977 & $-$70.118975795     \\
%%%$1j$ &   & 17.0000999865 & 115.626196037  & $4p$ &   & 12.970915034 & $-$67.475627239     \\
%%%$5s$ & 8 & 19.0000998250 & 113.148668458  & $1k$ & 8 & 14.092943322 & $-$72.356271359     \\
%%%$4d$ &   & 19.0000999599 & 114.791657698  & $2i$ &   & 14.343346418 & $-$71.000729286     \\
%%%$3g$ &   & 19.0000999776 & 116.618120456  & $3g$ &   & 14.561747154 & $-$69.261644312     \\
%%%$2i$ &   & 19.0000999844 & 117.404327590  & $4d$ &   & 14.722291210 & $-$66.986638351     \\
%%%$1k$ &   & 19.0000999881 & 117.797775019  & $5s$ &   & 14.797811479 & $-$63.950264091     \\
$5p$ & 9 & 21.0000999336 & 115.948021680  & $1l$ & 9 & 15.851969149 & $-$70.958113492     \\
$4f$ &   & 21.0000999712 & 117.926235639  & $2j$ &   & 16.091742759 & $-$69.765978162     \\
$3h$ &   & 21.0000999816 & 119.098277713  & $3h$ &   & 16.309144680 & $-$68.250662394     \\
$2j$ &   & 21.0000999864 & 119.632225052  & $4f$ &   & 16.483741176 & $-$66.287773320     \\
$1l$ &   & 21.0000999893 & 119.930276951  & $5p$ &   & 16.592021371 & $-$63.686362266     \\
\end{tabular}
\end{ruledtabular}
\footnotetext[1]{First nine levels are: 64.825083, 89.123452, 94.875967, 100.703996, 101.225824, 
105.507579, 103.184601, 105.945406, 108.527834, ref.~\cite{scherrer88}.}
\footnotetext[2]{First nine levels are: 64.740, 89.040, 93.360, 100.726, 101.108, 105.496, 99.440, 
106.030, 108.528, ref.~\cite{varshni87}.}
\footnotetext[3]{First nine levels are: 1.5604, 2.8614, 4.2814, 4.5098, 5.7882, 6.0512, 7.3612, 
7.6364, 7.8164, ref.~\cite{varshni87}.}
\end{table}
\endgroup

Table IV collects eigenvalues for four sets of coupling parameters; two for (+)ve and ($-$)ve 
$\lambda$ each. This table serves three purposes: (i) give energies for $\lambda <$0, for which no 
calculations have been made except those in \cite{varshni87}, (ii) give similar results for 
$\lambda > 0$, for which, as already mentioned earlier, accurate estimates, especially for higher energy 
levels, are quite scarce, and (iii) make a comparative study to discern the energy orderings in (+)ve 
and ($-$)ve $\lambda$ regimes side by side. To accomplish this, we systematically calculated all the 
30 energy levels of $n$ ranging from 0 to 9 for forty-five sets of ($g$,$\lambda$), with $g$ values 
0.1,1,10,100,1000 while the latter as $-$100,$-$10,$-$1,$-1$0.1,0.1,1,10,100,1000. Only seventeen of 
these states (corresponding to $n=0-5,9$) are presented here, which is sufficient to support the conclusions 
drawn herein. Note that in this table, we follow spectroscopic notation, i.e., the levels are labeled as 
$n_r$+1 and $l$ values. So $n_r$=3, $l$=2 signifies $4d$ level and so on. Out of all these 
($g$,$\lambda$) pairs, four are presented for reasons to become clear in the following. It may be 
mentioned that in \cite{varshni87,scherrer88}, energies up to $n$=4 leading to nine combinations of 
$n_r$,$l$ ($1s$--$1g$) were given for (+)ve $\lambda$, although in a different context (see later). 
For (+)ve $\lambda$, we find that the ordering given in column 1 holds true for twenty-three sets out of 
twenty-five, 
excepting (1,100) and (10,1000). That is why (10,1000) is picked up. The first instance of such a 
violation is encountered for between the levels ($1f$, $3s$) for $\lambda > 0$, and ($2s$, $1f$) for
$\lambda < 0$. The first nine states of the set (10,1000) reported in \cite{varshni87,scherrer88} 
interestingly show this feature qualitatively even though present values are far superior to those. 
Thereafter this ordering is not followed in many occasions as can be seen from the table. The 
(1000,0.1) set, besides being a representative of the ($g,\lambda$) set obeying the most observed 
sequence, also numerically illustrates that as $g$ increases, eigenvalues approach the harmonic 
oscillator values and in the limit of $\lambda\! \rightarrow$0, all the levels of a particular $n$ 
tend to be degenerate. Similar observations also hold for (1000,$-$0.1), except that now the 
eigenvalues approach harmonic oscillator values from below. Out of twenty ($g$,$\lambda$) pairs with 
($-$)ve $\lambda$, the ordering in column 5 is found to be valid for seventeen sets while the sets 
(0.1$-$10), 
(0.1,$-$100), (1,$-$100) do not follow this. So (0.1,$-$1) is a representative of those seventeen sets 
and also in this case first nine levels have been reported earlier in the literature (given in footnote). 
The (1,$-$100) set does not follow the same trend as (0.1, $-$1) and is given in last column. We note 
that the ``usual'' or mostly observed ordering in these two cases of (+)ve and ($-$)ve $\lambda$ for a 
given $n$ are mirror images of each other (or reversed). For example, when $n$=9, the energy 
ordering is $5p\!<\!4f\!<\!3h\!<\!2j\!<\!1l\!$ for (+)ve $\lambda$, while for ($-$)ve $\lambda$ it is 
$1l\!<\!2j\!<\!3h\!<\!4f\!<\!5p\!$.

\begin{figure}
\centering
\begin{minipage}[t]{0.4\textwidth}\centering
\includegraphics[scale=0.38]{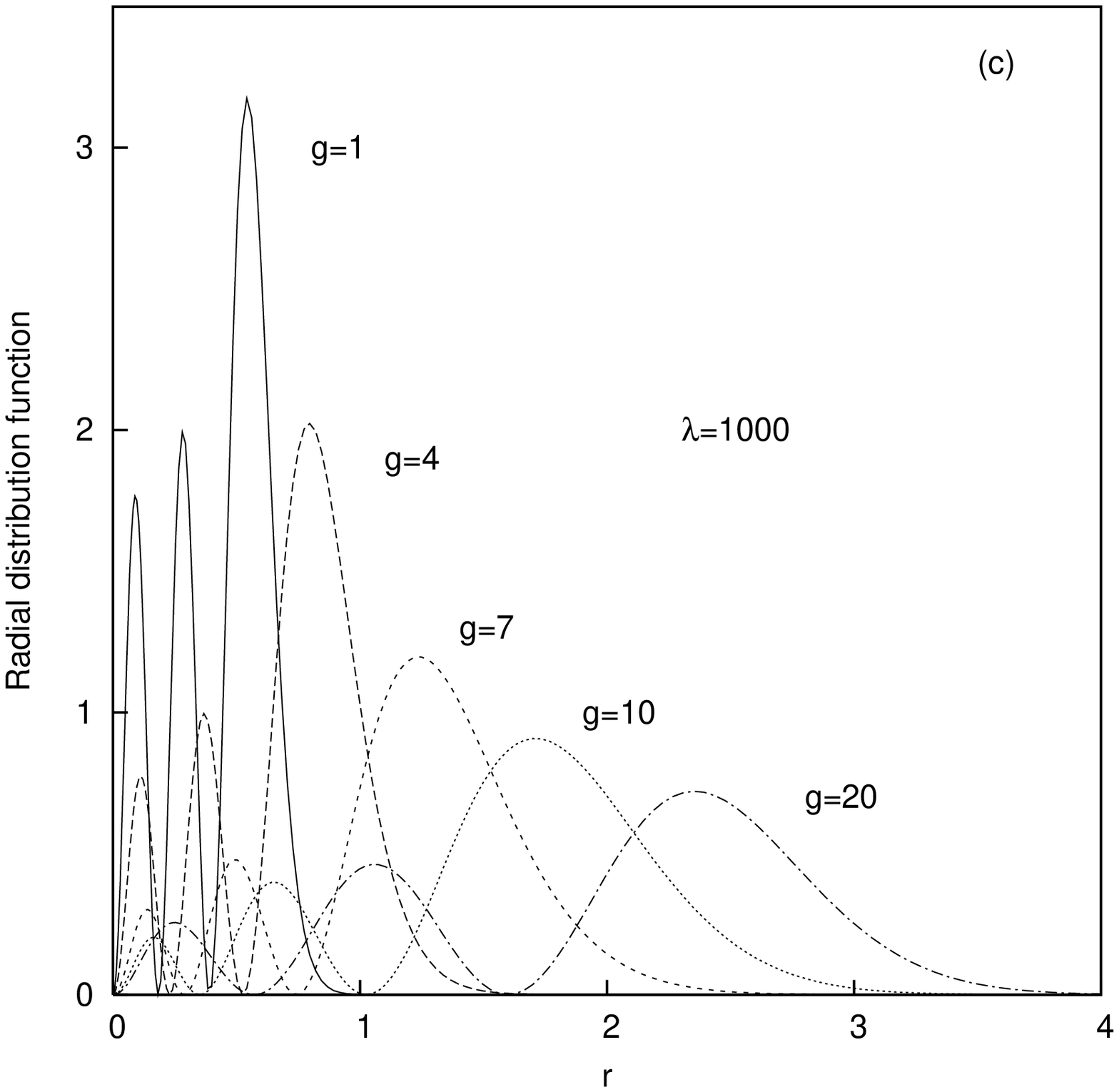}
\end{minipage}
\hspace{0.10in}
\begin{minipage}[t]{0.35\textwidth}\centering
\includegraphics[scale=0.38]{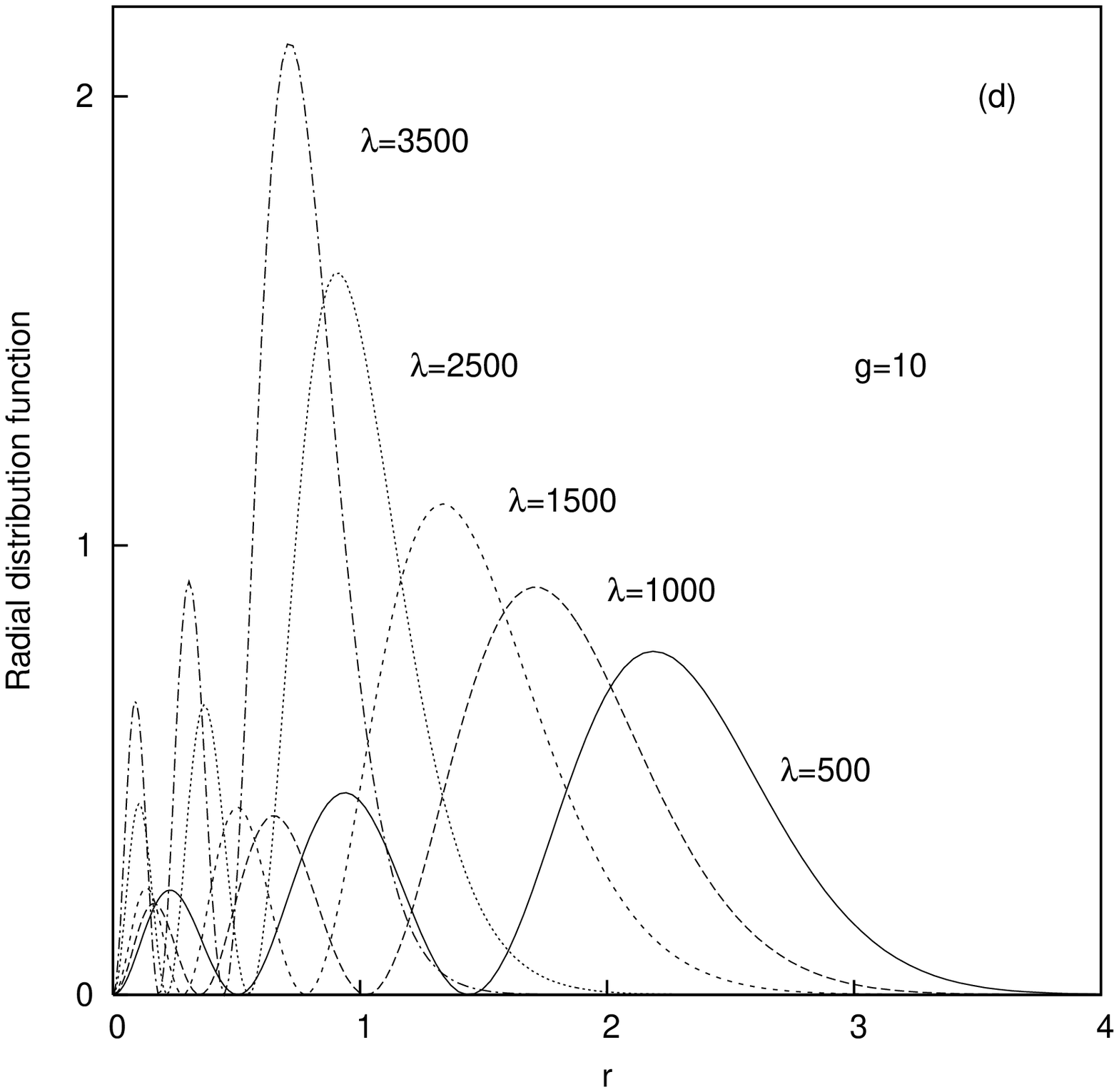}
\end{minipage}
\\[20pt]
\begin{minipage}[b]{0.4\textwidth}\centering
\includegraphics[scale=0.38]{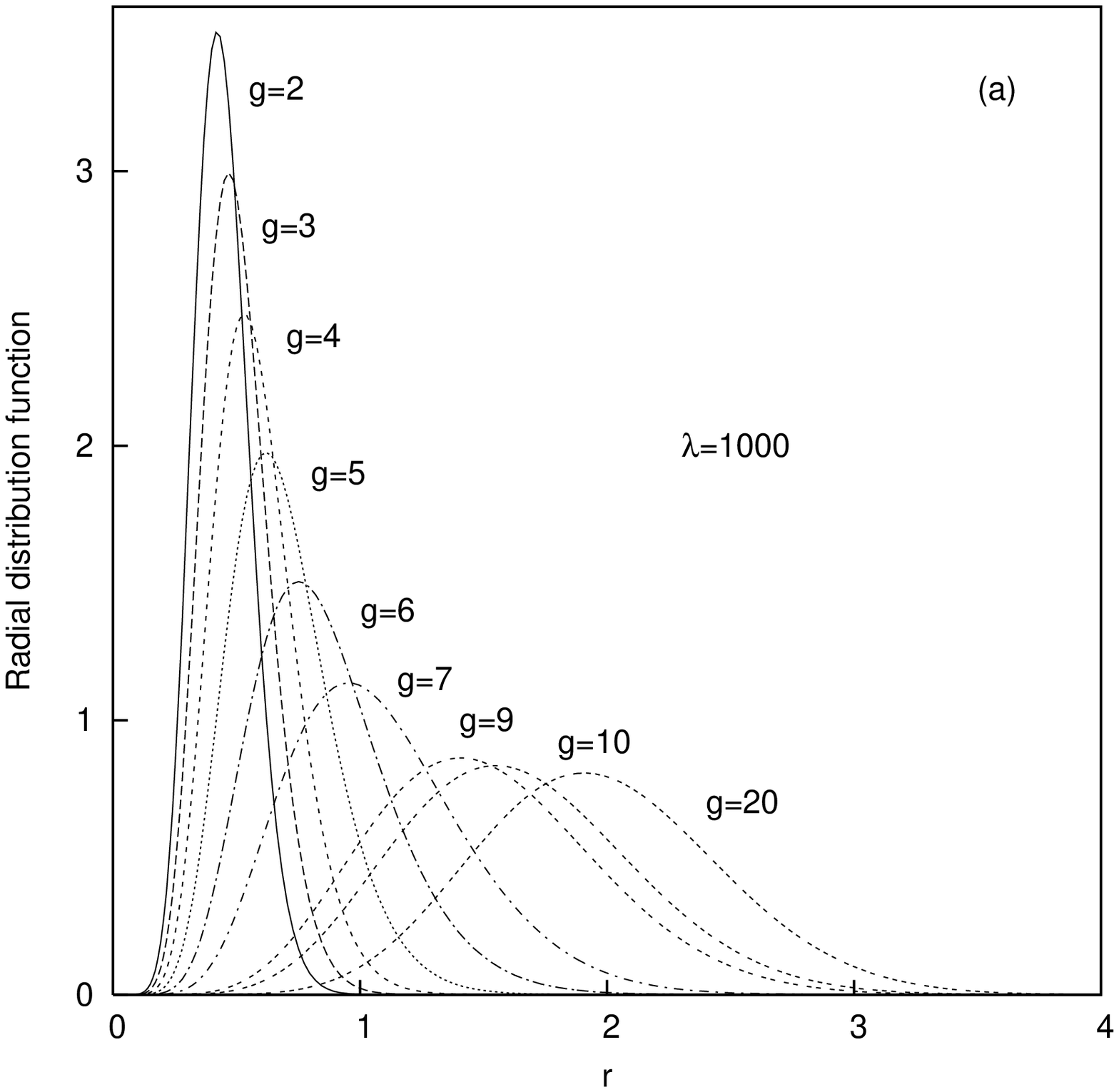}
\end{minipage}
\hspace{0.10in}
\begin{minipage}[b]{0.35\textwidth}\centering
\includegraphics[scale=0.38]{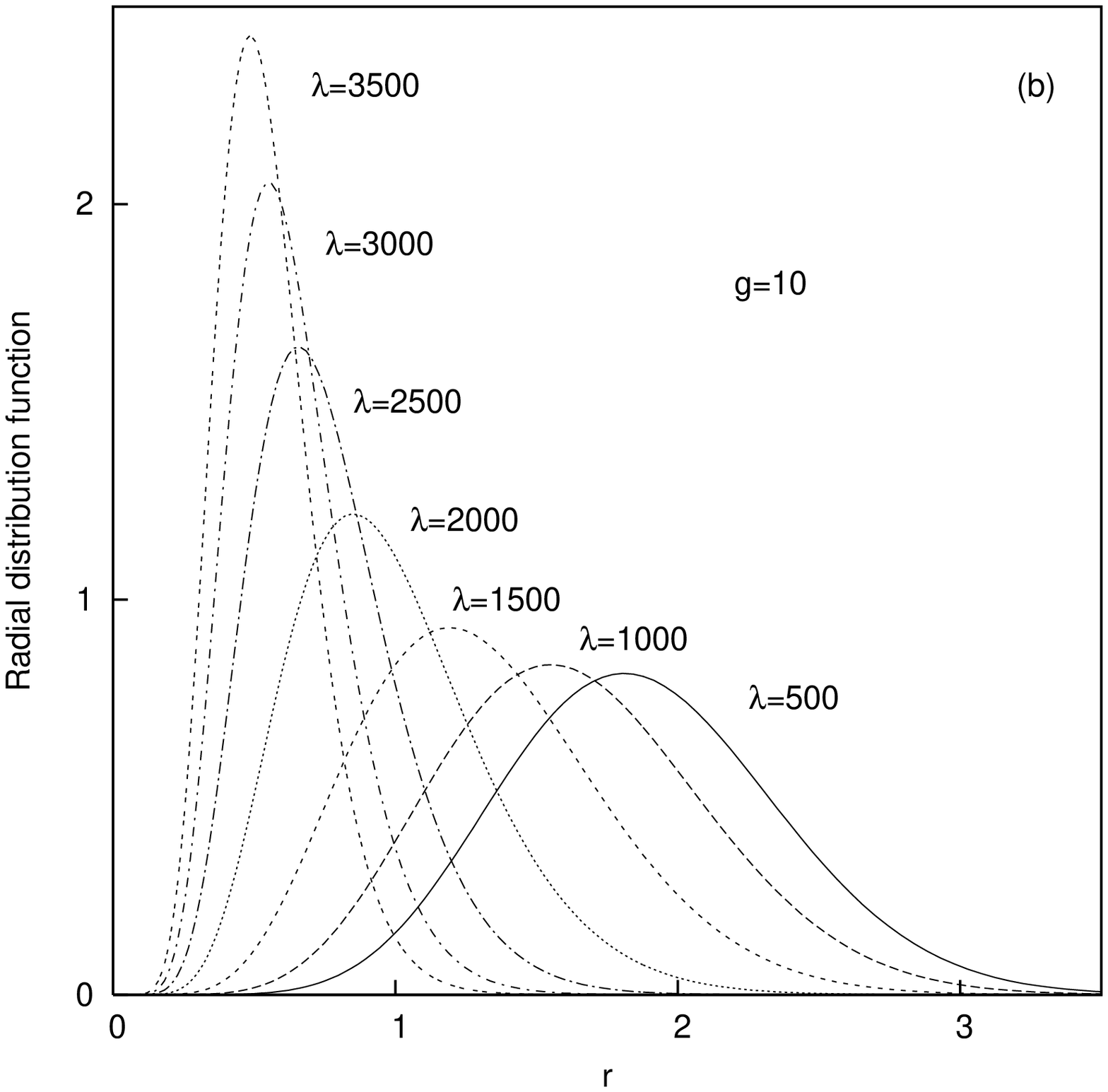}
\end{minipage}
\caption[optional]{Variation of the radial probability distribution functions, $|rR_{n,\ell}|^2$, 
for the $1f$ (bottom panel) and $3s$ (top panel) states of the 3D NPO. For (a) and (c), 
$\lambda=1000$; for (b) and (d), g=10. See text for details.}
\end{figure}

Now Fig. 2 displays the changes in radial distribution for the first two states ($1f$ and $3s$) in the 
vicinity of $g$ and $\lambda$ values, where the energy ordering is broken. The bottom panel ((a), (b)) 
corresponds to $1f$ state, while the top panel ((c), (d)) represents $3s$ state. (a) and (c) show
the density changes for a fixed $\lambda$ (1000) while varying $g$; (c) and (d) show the same for a 
fixed $g$ (10) with respect to variations in $\lambda$. For both the states, we note that, for a 
fixed $\lambda$, with an increase in $g$, the peak positions shift to the higher values of $r$, the
peak height decreases and the distribution broadens. For a fixed $g$, with an increase in $\lambda$,
correspondingly opposite trend is observed in (b) and (d) for both the states.  

\begingroup
\squeezetable
\begin{table}
\caption {\label{tab:table5}Variation of first twelve splittings ($\Delta \mathrm{E}$) of 3D NPO 
occurring between certain adjacent levels of $n$=2--7 with respect to $\lambda$ (left), and $g$ 
(right). Parentheses enclose the respective ($g, \lambda$) pairs. See text for details.}
\begin{ruledtabular}
\begin{tabular}{cccc|c|cccc}
  No. & (0.1,$-$0.1) & (0.1,$-$1) & (0.1,$-$10) & (0.1,$-$100) & (1,$-$100) & (10,$-$100) & (100,$-$100) & (1000,$-$100) \\
\hline
 1  & 0.014016 &  0.212252  & 3.008518  & 3.727614   & 3.124814  & 0.528414   & 0.011480  & 0.000144   \\
 2  & 0.018938 &  0.254371  & 2.828010  & 3.683210   & 2.719405  & 0.252782   & 0.003626  & 0.000038   \\
 3  & 0.021770 &  0.269873  & 2.657399  & 3.638930   & 2.358536  & 0.142225   & 0.001742  & 0.000018   \\
 4  & 0.010083 &  0.136933  & 2.976420  & 3.727042   & 3.097690  & 0.446619   & 0.010609  & 0.000140   \\
 5  & 0.023245 &  0.272365  & 2.497516  & 3.594793   & 2.044953  & 0.089829   & 0.001026  & 0.000010   \\
 6  & 0.014199 &  0.180442  & 2.786802  & 3.682481   & 2.682610  & 0.232326   & 0.003561  & 0.000038   \\
 7  & 0.023833 &  0.268071  & 2.348674  & 3.550816   & 1.776512  & 0.061534   & 0.000678  & 0.000001   \\
 8  & 0.016888 &  0.203685  & 2.609170  & 3.638047   & 2.317747  & 0.135666   & 0.001730  & 0.000018   \\
 9  & 0.007716 &  0.098492  & 2.940544  & 3.726466   & 3.068309  & 0.384311   & 0.009937  & 0.000137   \\
 10 & 0.023845 &  0.260169  & 2.210768  & 3.507016   & 1.548542  & 0.044682   & 0.000482  & 0.000005   \\
 11 & 0.018550 &  0.214770  & 2.444401  & 3.593757   & 2.004532  & 0.087238   & 0.001023  & 0.000010   \\
 12 & 0.011165 &  0.136694  & 2.740799  & 3.681748   & 2.643349  & 0.214832   & 0.003500  & 0.000038   \\
\end{tabular}
\end{ruledtabular}
\end{table}
\endgroup

It is well-known that for a 3D harmonic oscillator $n_r$ and $l$ quantum numbers satisfying 
$n\!=\!2n_r+l$ are degenerate. Thus for $n$=4, ($n_r$,$l$) pairs having values (2,0), (1,2) and 
(0,4) are degenerate, etc. For non-zero $\lambda$, such degeneracies vanish and these are 
conveniently analyzed through the respective level spacings, 
$\Delta \mathrm{E} \! =\mathrm{E}_{n_r,l}-\mathrm{E}_{n_r',l'}$. Quite detailed studies were 
conducted for first four ($n\!\!\le$4) such splittings of positive $\lambda$ \cite{varshni87,scherrer88}. We 
found similar qualitative features for those in \cite{scherrer88} and thereby not repeated here. 
However, to our knowledge, no such attempts have been made as yet for $\lambda\!<$0. Table V gives 
variations of all 
the first twelve such splittings possible between certain adjacent levels of $n$=2--7 with respect to 
the interaction parameters. First two splittings $\mathrm{E}_{1,0}\!-\!\mathrm{E}_{0,2}$ and 
$\mathrm{E}_{1,1}\!-\!\mathrm{E}_{0,3}$ belong to $n$=2,3 respectively, while the last three 
$\mathrm{E}_{1,5}\!-\!\mathrm{E}_{0,7}$, $\mathrm{E}_{2,3}\!-\!\mathrm{E}_{1,5}$, 
$\mathrm{E}_{3,1}\!-\!\mathrm{E}_{2,3}$ correspond to $n$=7. In columns 2--5, $g$ is kept fixed at 
0.1, while $\lambda$ varied from $-$0.1 to $-$100. Clearly all these splittings tend to increase as 
$|\lambda|$ increases; a trend as also observed for $\lambda\!\!>$0 \cite{scherrer88}. 
Furthermore, according to them the first four splittings, i.e., the reverse of ours 
($\mathrm{E}_{0,2}\!-\!\mathrm{E}_{1,0}$, $\mathrm{E}_{0,3}\!-\!\mathrm{E}_{1,1}$, 
$\mathrm{E}_{1,2}\!-\!\mathrm{E}_{2,0}$, $\mathrm{E}_{0,4}\!-\!\mathrm{E}_{1,2}$) increases 
continuously with $\lambda$, eventually approaching a constant value in the limit of 
$\lambda \!\rightarrow \!\infty$. However they considered $\lambda\!>\!10^4$ and more elaborate 
calculations would be required to confirm whether such a trend will also occur for $\lambda\!<$0. 
Similar trend is observed for other $g$ as well as the remaining eight splittings (not given) occurring 
between certain permissible levels within $n$=8 and 9. Finally we examine the changes in spacings 
with respect to $g$ (from 0.1--1000), keeping $\lambda$ fixed at $-$100 (in columns 5--9). Thus the 
entries belonging to (0.1,$-$100) in column 5 are common to both the variations and hence enclosed 
in box with border from both sides. Clearly they tend to disappear in the limit of large $g$, in 
agreement with a conclusion found in \cite{varshni87,scherrer88} for positive $\lambda$. Once again, 
this has been found to be numerically true for the rest eight splittings occurring between higher 
levels (not presented in the table) as well as for other values of $\lambda$ ($-$0.1,$-$1,$-$10). 

Before passing, it may be noted that, in the last decade, the power series solution method (originally 
proposed by \cite{campoy86, palma87}) and some of its variants have been very successful in producing 
quite accurate eigenvalues, radial 
expectation values, etc., of a number of systems such as the confined 3D isotropic harmonic oscillator, 
2D H atom confined in a circle with impenetrable walls, 3D H atom confined in an impenetrable sphere
\cite{campoy02,aquino05,aquino07}. Very recently, highly accurate eigenvalues have also been reported 
\cite{aquino07} by means of a formal solution of the confluent hypergeometric function for the 
confined H atom. It may be interesting to employ such methods in the context of the NPOs.

\section{Conclusion}
Quantum mechanics has nowadays widely spread applications almost everywhere in contemporary science 
and technology. However, exactly solvable quantum systems are very scarce. Nearly exact numerical 
solutions of quantum-mechanical problems is of paramount interest in this vein, and also for the 
merits of the theory itself. This work offers nearly exact solutions for a specific class of quantum 
potentials, the non-polynomial oscillator class, that has broad applications in various quantum and 
quantum-field models. Accurate bound states of the 3D NPO are calculated and considered in detail. 
While several high-quality results have been published for the 1D NPO, the data for 3D NPO is quite 
limited. Reliable variational bounds have become available only very recently. Moreover, the 
$\lambda\!\!<$0 region is explored very little. For both positive and negative $\lambda$, as well as 
for far high-lying states, the present GPS calculations match very well with the best literature results. 
All levels in the range of $n$=0--9 (comprising thirty overall) are investigated for a wide range 
of the coupling parameters. This study surpasses all the existing results except the finite-difference 
data of ref. \cite{witwit91}, which offers similar accuracy.

Energy and radial density distribution variations are followed as functions of $g$ and $\lambda$. An 
attempt is made to identify the energy orderings for $\lambda\!<$0 and $\lambda\!>$0 separately. 
Although there seems to be a qualitative pattern, several deviations of oscillatory character to it
are observed. Finally all the twenty energy differences between certain admissible levels for $n$=0--9 
are examined for the first time as functions of the parameters. The method employed here is simple, 
easily applies to a wide variety of strong and weak interactions and performs very well for low and 
higher excitations. We hope that our results may serve as benchmarks for these potentials. 

\begin{acknowledgments}
AKR thanks professors D.~Neuhauser and S. I. Chu for support and useful discussions. He acknowledges
the warm hospitality provided by the Univ. of California, Los angeles, CA, USA. EP gratefully
acknowledges Q-Chem Inc., for support. We thank the anonymous referee for valuable and constructive
comments, which helped to improve the manuscript.
\end{acknowledgments}

\end{document}